\documentclass[thmsa,a4paper,notitlepage,11pt]{article}
\usepackage{amssymb}
\usepackage{amsmath}
\usepackage{graphicx}
\usepackage[compress,adjust,nospace]{cite}
\usepackage{bibmods}

\input{tcilatex}

\begin{document}

\title{Symmetries and Conservation Laws for the $2D$ Ricci Flow Model }
\author{Rodica Cimpoiasu \\
%EndAName
University of Craiova, 13 A.I.Cuza, 200585 Craiova, Romania,\\
e-mail address: rodicimp@yahoo.com}
\date{}
\maketitle

\begin{abstract}
The paper aims to study the connection between symmetries and conservation
laws for the $2D$ Ricci flow model. The procedure starts by obtaining a set
of multipliers which generates conservation laws. Then, using a general
relation which connects symmetries and conservation laws for whatever
dynamical system, one determines symmetries related to a chosen multiplier.
On this basis, new similarity solutions of the model, not yet discussed in
literature, are highlighted.


\end{abstract}

\section{\protect\smallskip Introduction}

The concepts of symmetry, invariants and conservation laws are fundamental
in the study of dynamical systems, providing a clear connection between
equations of motion and their solutions. There are many reasons for
computing the symmetries and the conservation laws corresponding to systems
of which the evolution is described by differential equations. In the recent
years, a remarkable number of mathematical models occurring in various
research domains have been studied from the point of view of the symmetry
groups theory \cite{Ames, Blu, Hyd}. The most suitable technique for finding
classes of analytical solutions, the so-called Lie symmetry method,
investigates integrability starting from the invariance of evolutionary
equations under some linear transformations of the variables which define
the so-called Lie group of symmetries. Conservation laws proceed from the
conservation of physical quantities: linear momentum, mass,energy, electric
charge, etc. Furthermore, conservation laws are applied in the study of
partial differential equations, as for example in: (i) test of complete
integrability and application of Inverse Scattering Transform, (ii) study of
quantitative and qualitative properties for pdes (Hamiltonian structure,
recursion operators) and (iii) development of numerical methods such as
finite element methods \cite{Rand, Berna}.

There are many interesting results on the correspondence between symmetries
and conservation laws. As examples, an identity \cite{AB} which does not
depend on the use of a Lagrangian and provides a relationship between
symmetries and conservation laws for self-adjoint differential equations is
derived; a direct link between the components of a conserved vector of an
arbitrary partial differential equation and the Lie-B\"{a}cklund symmetry
generator of the equation, which is associated with the conserved vector's
components is obtained \cite{Mahomed}.

The aim of this paper is to tackle the connection between symmetries and
conservation laws using the results mentioned above and the property of the
Euler differential operator to annihilate whatever expression having the
mathematical form of a divergence. By applying the Euler operator on a
combination of evolutionary equations, one can determine a set of
multipliers generating conservation laws. We will apply this procedure for
the $2D$ Ricci flow equation, a very interesting model coming from the
gravity theory. The Lie symmetry problem and the invariant solutions of this
model have been already discussed \cite{Ricci}. Using the procedure
mentioned above, new symmetries and solutions will be highlighted.

The paper is organized as follows: after this introductive section, the
problem of constructing a conserved vector will be analyzed in the second
section, by means of the conservation law multipliers' method and also by
tacking into account the relations which impose an intimate connection
between symmetry generators and conserved vectors. In the third section,
these methods will be applied to the $2D$ Ricci flow model. A set of
conservation laws corresponding to a multiplier $\Lambda (x,t,y,U)$ that
does not depend on the derivatives of the dependent variable $U$, the
associated Lie symmetry operators and some new invariant solutions, will be
obtained. Some concluding remarks will end the paper.

\section{ Connection between symmetries and conservation laws}

For each partial differential equation (pde) or for each pde system there is
a local group of transformations (called the symmetry group) that acts on
the space of its independent and dependent variables, with the property that
it maps the set of all analytical solutions to itself, and so it leaves the
form of the equation (or system) unchanged. The widely applicable method to
find the symmetry group associated with a pde (or pde system) is called the 
\textit{classical Lie method}. Consequently, the knowledge of Lie point
symmetries allows us to construct the group-invariant solutions. Two
solutions should be equivalent if there whould be a symmetry transformation
that could transform the one into the other. Moreover, new solutions might
be obtained from the known ones: by applying the symmetry group to a known
solution, a family of new solutions would be created.

Let us consider a $n$-th order partial differential system:%
\begin{equation}
\Delta ^{\nu }[x,u(x),u^{(n)}(x)]=0,\text{ }\nu =\overline{1,q}  \label{2.1}
\end{equation}%
where $x\equiv \{x^{i},$ $i=\overline{1,p}\}\subset R^{p}$ represent
independent variables, while $u\equiv \{u^{\alpha },\alpha =\overline{1,q}%
\}\subset R^{q}$ dependent ones. The notation $u^{(n)}$designates the set of
partial derivatives of $u$ with respect to $x,$ up to the $n$-th order.

Let us consider the infinitesimal symmetry operator with the general form: 
\begin{equation}
X=\dsum\limits_{i=1}^{p}\xi ^{i}(x,u)\frac{\partial }{\partial x^{i}}%
+\dsum\limits_{\alpha =1}^{q}\phi _{\alpha }(x,u)\frac{\partial }{\partial
u^{\alpha }}  \label{2.2}
\end{equation}

In order to explicit the above mentioned property of classical symmetries,
the Lie method \cite{Olver} applied the criterion of infinitesimal
invariance of the system (\ref{2.1}) to the action of the operator (\ref{2.2}%
). More precisely, the following condition ought to be imposed:%
\begin{equation}
X^{(n)}(\Delta ^{\nu })\left\vert _{\Delta =0}\right. =0  \label{2.3}
\end{equation}%
where $X^{(n)}$ represents the extension of $n-$th order of the Lie symmetry
generator (\ref{2.2}).

A conservation law for partial differential equations is a divergence
expression which vanishes for the solutions of the pde system. The origin of
conservation laws comes from physical principles such as the conservation of
mass, momentum and energy. Although there is a well-known systematic method 
\cite{Noether} able to find conservation laws for variational pdes, the
applicability of Noether's method is limited by the fact that many
interesting pde systems are not variational. A systematic procedure able to
find conservation laws, called the direct method, has been developed \cite%
{Anco, Bluman}. The direct method consists of \ two main steps :

$(i)$ to determine a set of conservation law multipliers so that a linear
combination of the pdes with the conservation law multipliers should create
a divergence expression. For our given differential system (\ref{2.1}) one
could define a set of conservation law multipliers $\{\Lambda _{\nu }:$ $M$ $%
\times $ $S^{(n)}\rightarrow R\},$ $\nu =\overline{1,q},$ if smooth
functions $\{P_{i}:$ $M$ $\times $ $S^{(n)}\rightarrow R\},$ $i=\overline{1,p%
}$ should exist such that everywhere on $M$ $\times $ $S^{(n)}$ we could
have:%
\begin{equation}
\Lambda _{\nu }[x,U(x),U^{(n)}(x)]\Delta ^{\nu
}[x,U(x),U^{(n)}(x)]=D_{i}P^{i}[x,U(x),U^{(n)}(x)]  \label{2.4}
\end{equation}%
It is known that conservation law multipliers could be found using the
method \cite{Bluman} of the Euler operator. Thereby, we can solve the
following system of pdes:%
\begin{equation}
E_{\rho }\left[ \Lambda _{\nu }[x,U(x),U^{(n)}(x)]\Delta ^{\nu
}[x,U(x),U^{(n)}(x)]\right] =0,\text{ }\forall \text{ }\rho =\overline{1,q}
\label{2.5}
\end{equation}%
for the unknown functions $\{\Lambda _{\nu },$ $\nu =\overline{1,q}\}.$

$(ii)$ ones having determined a set of conservation law multipliers, to find
the corresponding fluxes in order to obtain the conservation law:

\begin{eqnarray}
D_{i}P^{i}[x,U(x),U^{(n)}(x)]\text{ }_{\left\vert U(x)=u(x)\right. }
&=&\Lambda _{\nu }[x,U(x),U^{(n)}(x)]\Delta ^{\nu
}[x,U(x),U^{(n)}(x)]_{\left\vert U(x)=u(x)\right. }=  \notag \\
\Lambda _{\nu }[x,u(x),u^{(n)}(x)]\Delta ^{\nu }[x,u(x),u^{(n)}(x)] &=&0
\label{2.6}
\end{eqnarray}%
\textbf{\ }

The key property of this set of conservation law multipliers is that their
existence implies the existence of conservation laws. Conversely, it turns
out that for non-degenerate pde systems, every conservation law up to
equivalence must arise from a set of conservation law multipliers.\textbf{\ }%
Generally, the correspondence between sets of conservation law multipliers
and equivalent conservation laws could be many-to-one. However, if a pde
system admits a Cauchy-Kolvalevskaya form \cite{CK} and the sets of
conservation law multipliers satisfy some mild conditions, then there is a
one-to-one correspondence between each set of conservation law multipliers
and each set of equivalent conservation laws.

Further, a relation \cite{Mahomed} was derived between the symmetry operator 
$X$ and the components of a conserved vector $P_{i}.$ It has the form:%
\begin{equation}
X^{(r)}(P^{i})+P^{i}D_{k}(\xi ^{k})-P^{k}D_{k}(\xi ^{i})=0,\text{ }i=%
\overline{1,p}  \label{2.7}
\end{equation}%
where $X$ is applied in the $r-$th extended form.

The conditions (\ref{2.7}) with $X$ known, joined to the conservation law $%
D_{i}P^{i}=0,$ might be viewed as a system of linear partial differential
equations which could be solved for the components $P^{i},$ $i=\overline{1,p}
$ of the conserved vector. Yet, (\ref{2.7}) could be used as well to obtain
the symmetry operators $X$ associated with a given conserved vector. In the
next section, the latter way will be applied.

\section{Application to the 2D Ricci flow model}

\subsection{Multipliers of the model}

In this section, the relationship between symmetries and conservation laws
will be applied to the $2D$ Ricci flow model. Its evolutionary equation,
bearing large consequences in gravity theory, has the form:%
\begin{equation}
u_{t}=\frac{u_{xy}}{u}-\frac{u_{x}u_{y}}{u^{2}}  \label{3.1}
\end{equation}

It is a well-known equation which has been studied as a continuum limit of
the Toda-type equation. Among the main results concerning (\ref{3.1}) one
may mention: $(i)$\textit{\ }it could be obtained as a particular case of
the $3D$ Ricci flow equation which accepts a Killing vector;$(ii)$ by
linearization, it presents various classes \cite{adi} of solutions, these
depending on the \textquotedblright sector\textquotedblright\ where it is
defined; $(iii)$ an effective study of its Lie symmetries and invariant
solutions was performed \cite{Ricci}.

Now, the first task will be to obtain conservation laws for the previous
model, using the multipliers' method. A multiplier $\Lambda $ of equation (%
\ref{3.1}) has the property :%
\begin{equation}
\Lambda \left[ U_{t}-\frac{U_{xy}}{U}+\frac{U_{x}U_{y}}{U^{2}}\right]
=D_{i}P^{i}[t,x,y,U,U^{(2)}]  \label{3.2}
\end{equation}%
for all functions $U(t,x,y)$, not only for the solutions $u(t,x,y)$ of (\ref%
{3.1}).

Let us consider multipliers of the form $\Lambda =\Lambda (t,x,y,U).$
Multipliers which depend on the first order and higher order partial
derivatives of $U$ could also be considered. Yet, in two dimensions,
calculations rapidly become more complicated. Therefore, computer assisted
calculations may lead to further conservation laws.

The right hand side of (\ref{3.2}) is a divergence expression. Hence, this
expression should vanish identically when applying the Euler operator which,
in two dimensions, takes the form:%
\begin{equation}
E_{U}=\frac{\partial }{\partial U}-D_{t}\frac{\partial }{\partial U_{t}}%
-D_{x}\frac{\partial }{\partial U_{x}}-D_{y}\frac{\partial }{\partial U_{y}}%
+D_{x}^{2}\frac{\partial }{\partial U_{2x}}+D_{y}^{2}\frac{\partial }{%
\partial U_{2y}}+D_{x}D_{y}\frac{\partial }{\partial U_{xy}}+...  \label{3.3}
\end{equation}%
The determining equation for the multiplier $\Lambda (t,x,y,U)$ is:%
\begin{equation}
E_{U}\left[ \Lambda \left( U_{t}-\frac{U_{xy}}{U}+\frac{U_{x}U_{y}}{U^{2}}%
\right) \right] =0  \label{3.4}
\end{equation}%
The expansion of (\ref{3.4}) yields the following pde:%
\begin{equation}
-\frac{2\Lambda _{U}}{U}U_{xy}+\left( \frac{\Lambda _{U}}{U^{2}}-\frac{%
\Lambda _{2U}}{U}\right) U_{x}U_{y}-\Lambda _{yU}U_{x}-\Lambda
_{xU}U_{y}-\left( \Lambda _{t}+\frac{\Lambda _{xy}}{U}\right) =0  \label{3.5}
\end{equation}%
Because (\ref{3.5}) is satisfied for all functions $U(t,x,y)$ and because
the multiplier $\Lambda $ is chosen as not to depend on derivatives of $U$,
the coefficient functions of various derivatives of $U$ should vanish.
Thereby, the determining system for $\Lambda (t,x,y,U)$ is generated:%
\begin{equation}
\Lambda _{U}=\Lambda _{t}=\Lambda _{xy}=0  \label{3.6}
\end{equation}%
By solving the system (\ref{3.6}), one obtains the general multiplier: 
\begin{equation}
\Lambda =f(x)+g(y),\text{ }\forall f(x),\text{ }\forall g(y)  \label{3.7}
\end{equation}

\subsection{Conservation laws and symmetries}

The first aim of this section is to determine the conserved vector
associated with the multiplier (\ref{3.7}). From (\ref{3.2}) and (\ref{3.7})
and by performing elementary manipulations, one obtains:%
\begin{equation}
\lbrack f(x)+g(y)]\left[ U_{t}-\frac{U_{xy}}{U}+\frac{U_{x}U_{y}}{U^{2}}%
\right] =D_{t}\left[ (f(x)+g(y))U\right] +D_{x}\left[ -g(y)\frac{U_{y}}{U}%
\right] +D_{y}\left[ -f(x)\frac{U_{x}}{U}\right] =D_{i}P^{i}  \label{3.8}
\end{equation}%
Consequently, when $U(t,x,y)$ is a solution of (\ref{3.1}), the previous
expression leads to the following set of conservation laws:%
\begin{equation}
D_{t}\left[ (f(x)+g(y))u\right] +D_{x}\left[ -g(y)\frac{u_{y}}{u}\right]
+D_{y}\left[ -f(x)\frac{u_{x}}{u}\right] =0,\text{ }\forall f(x),\text{ }%
\forall g(y)  \label{3.9}
\end{equation}%
As a result, one obtains that whatever conserved vector of the Ricci flow
model (\ref{3.1}) with multiplier of the form (\ref{3.7}), should have the
components:%
\begin{equation}
P^{1}=(f(x)+g(y))u,\text{ }P^{2}=-g(y)\frac{u_{y}}{u},\text{ }P^{3}=-f(x)%
\frac{u_{x}}{u},  \label{3.10}
\end{equation}%
\qquad \qquad

The next aim is to illustrate how we might find the point symmetries
associated with the conservation law (\ref{3.9}). The starting point of
computations is clearly the relation (\ref{2.7}) in which we consider $P^{i}$
as known components of the conserved vector and $X$ as an unknown symmetry
operator. The symmetry conditions for (\ref{3.10}) are:%
\begin{eqnarray}
X^{(1)}(P^{1})+P^{1}D_{x}(\xi )+P^{1}D_{y}(\eta )-P^{2}D_{x}(\varphi
)-P^{3}D_{y}(\varphi ) &=&0  \notag \\
X^{(1)}(P^{2})+P^{2}D_{t}(\varphi )+P^{2}D_{y}(\eta )-P^{1}D_{t}(\xi
)-P^{3}D_{y}(\xi ) &=&0  \label{3.11} \\
X^{(1)}(P^{3})+P^{3}D_{t}(\varphi )+P^{3}D_{x}(\xi )-P^{1}D_{t}(\eta
)-P^{2}D_{x}(\eta ) &=&0  \notag
\end{eqnarray}%
where the symmetry operator $X$ is of type (\ref{2.2}) and has the first
extension:%
\begin{equation}
X^{(1)}(t,x,y,u)=\varphi \frac{\partial }{\partial t}+\xi \frac{\partial }{%
\partial x}+\eta \frac{\partial }{\partial y}+\phi \frac{\partial }{\partial
u}+\phi ^{t}\frac{\partial }{\partial u_{t}}+\phi ^{x}\frac{\partial }{%
\partial u_{x}}+\phi ^{y}\frac{\partial }{\partial u_{y}}  \label{3.12}
\end{equation}%
Coefficient functions $\phi ^{t},$ $\phi ^{x},$ $\phi ^{y}$ will be
calculated according to the symmetry theory \cite{Olver}.

Let us suppose that $\Lambda =$ $f(x)+g(y)\neq 0.$ The expansion of the
determining equations (\ref{3.11}) and separation by various monomials in
derivatives of $u,$ do generate the following differential system:%
\begin{eqnarray}
\varphi _{u} &=&0  \notag \\
\xi _{y} &=&0  \notag \\
\xi _{u} &=&0  \notag \\
\eta _{x} &=&0  \notag \\
\eta _{u} &=&0  \notag \\
\varphi _{x}g(y) &=&0  \notag \\
\varphi _{y}f(x) &=&0  \notag \\
u\left[ \xi \frac{\text{ }df(x)}{dx}+\eta \text{ }\frac{dg(y)}{dy}%
+(f(x)+g(y))(\xi _{x}+\eta _{y})\right] +[f(x)+g(y)]\phi &=&0  \label{3.13}
\\
u\eta \text{ }\frac{dg(y)}{dy}-g(y)[\phi -u\phi _{u}-u\varphi _{t}] &=&0 
\notag \\
u\xi \text{ }\frac{df(x)}{dx}-f(x)[\phi -u\phi _{u}-u\varphi _{t}] &=&0 
\notag \\
u^{2}[f(x)+g(y)]\xi _{t}+g(y)\phi _{y} &=&0  \notag \\
u^{2}[f(x)+g(y)]\eta _{t}+f(x)\phi _{x} &=&0  \notag
\end{eqnarray}%
with unknown functions $\varphi ,$ $\xi ,$ $\eta ,$ $\phi $.

Looking to the system (\ref{3.13}), we are able to distinguish two cases for
solving it. More precisely, the latter five equations of the system (\ref%
{3.13}) should be solved by taking into account the following situations:

\begin{equation}
Case(I):f(x)\neq 0,g(y)\neq 0\Rightarrow \varphi (t),\xi (t,x),\eta
(t,y),\phi (t,x,y,u)  \label{3.132}
\end{equation}

and 
\begin{eqnarray}
Case(II) &:&f(x)=0,g(y)\neq 0\Rightarrow \varphi (t,y),\xi (t,x),\eta
(t,y),\phi (t,x,y,u)\text{ or}  \notag \\
g(y) &=&0,f(x)\neq 0\Rightarrow \varphi (t,x),\xi (t,x),\eta (t,y),\phi
(t,x,y,u)  \label{3.133}
\end{eqnarray}

\subsection{Symmetries and invariant solutions for Case (I)}

Under the conditions (\ref{3.132}), through a computational method (Maple 10
program), six solutions could be obtained. They correspond to the following
Lie symmetry operators:

\bigskip (1) $\forall a_{1}=const.,$ $\forall f(x)\neq const.,$ $\forall
g(y)\neq const.$%
\begin{equation}
X_{1}^{(I)}=a_{1}\frac{\partial }{\partial t}  \label{3.14}
\end{equation}

\bigskip (2) $\forall b_{j}.=const.,$ $j=1,3,4,$ $\forall
g(y)=b_{2}=const.\neq 0,$ $\forall f(x)\neq const.$%
\begin{equation}
X_{2}^{(I)}=b_{3}\frac{\partial }{\partial t}+[-b_{1}y+b_{4}]\frac{\partial 
}{\partial y}+b_{1}u\frac{\partial }{\partial u}  \label{3.15}
\end{equation}

(3) $\forall p_{j}.=const.,$ $j=1,3,4,$ $\forall f(x)=p_{2}=const\neq 0,$ $%
\forall g(y)\neq const.$%
\begin{equation}
X_{3}^{(I)}=p_{3}\frac{\partial }{\partial t}+[-p_{1}x+p_{4}]\frac{\partial 
}{\partial x}+p_{1}u\frac{\partial }{\partial u}  \label{3.16}
\end{equation}

(4) $\forall s_{k}.=const.,$ $k=\overline{1,7},k\neq 2,3,$ $\forall
f(x)=s_{2}=const\neq 0,$ $\forall g(y)=s_{3}=const\neq 0$%
\begin{equation}
X_{4}^{(I)}=s_{4}\frac{\partial }{\partial t}+[s_{5}x+s_{6}]\frac{\partial }{%
\partial x}+[(-s_{5}-s_{1})y+s_{7}]\frac{\partial }{\partial y}+s_{1}u\frac{%
\partial }{\partial u}  \label{3.17}
\end{equation}

(5) $\forall $ $m_{i}=const.,$ $i=\overline{1,4},$ $f(x)=(x-m_{1}),$ $%
g(y)=[-2(y+m_{2})]^{(-1/2)}$%
\begin{eqnarray}
X_{5}^{(I)} &=&[m_{3}t+m_{4}]\frac{\partial }{\partial t}-m_{3}[x-m_{1}]%
\frac{\partial }{\partial x}+2m_{3}[y+m_{2}]\frac{\partial }{\partial y}= 
\notag \\
&&[m_{3}t+m_{4}]\frac{\partial }{\partial t}-m_{3}f(x)\frac{\partial }{%
\partial x}-\frac{m_{3}}{2}\left[ g(y)\right] ^{(-2)}\frac{\partial }{%
\partial y}  \label{3.18}
\end{eqnarray}

(6) $\forall $ $n_{i}=const.,$ $i=\overline{1,4},$ $f(x)=(x-n_{2}),$ $%
g(y)=(y-n_{3})$%
\begin{eqnarray}
X_{6}^{(I)} &=&\left[ \frac{n_{1}}{3}t+n_{4}\right] \frac{\partial }{%
\partial t}-\frac{n_{1}(x-n_{2})}{3}\frac{\partial }{\partial x}-\frac{%
n_{1}(y-n_{3})}{3}\frac{\partial }{\partial y}+n_{1}u\frac{\partial }{%
\partial u}=  \notag \\
&&\left[ \frac{n_{1}}{3}t+n_{4}\right] \frac{\partial }{\partial t}-\frac{%
n_{1}f(x)}{3}\frac{\partial }{\partial x}-\frac{n_{1}g(y)}{3}\frac{\partial 
}{\partial y}+n_{1}u\frac{\partial }{\partial u}  \label{3.19}
\end{eqnarray}%
It is important to remind some results \cite{Ricci}:

$(i)$ by applying the classical symmetry approach, the general Lie operator
for the Ricci flow model (\ref{3.1}) is obtained in the form:%
\begin{equation}
V=(c_{1}t+c_{2})\frac{\partial }{\partial t}+\xi (x)\frac{\partial }{%
\partial x}+\eta (y)\frac{\partial }{\partial y}+u\left[ c_{1}-\frac{d\xi (x)%
}{dx}-\frac{d\eta (y)}{dy}\right] \frac{\partial }{\partial u}  \label{3.20}
\end{equation}%
with $c_{1},$ $c_{2}$ arbitrary constants and $\xi (x),$ $\eta (y)$
arbitrary functions.

$(ii)$ in the linear sector of invariance (when $\xi (x),$ $\eta (y)$ are
considered as having linear forms), the Ricci flow model admits a $6-$%
parameters family of Lie operators which generates the following $6$
independent symmetry operators:%
\begin{eqnarray}
V_{1} &=&t\frac{\partial }{\partial t}+u\frac{\partial }{\partial u},\text{ }%
V_{2}=\frac{\partial }{\partial t},\text{ }V_{3}=\frac{\partial }{\partial x}%
,\text{ }V_{4}=\frac{\partial }{\partial y},\text{ }  \notag \\
V_{5} &=&x\frac{\partial }{\partial x}-u\frac{\partial }{\partial u},\text{ }%
V_{6}=y\frac{\partial }{\partial y}-u\frac{\partial }{\partial u}
\label{3.21}
\end{eqnarray}

\textit{Remark 1} : The Lie point symmetries associated to the conservation
law (\ref{3.9}), which have the expressions (\ref{3.14})-(\ref{3.19}) could
be expressed as a linear combination of the independent operators (\ref{3.21}%
). These expressions are:%
\begin{eqnarray}
X_{1}^{(I)} &=&a_{1}V_{2},\text{ }%
X_{2}^{(I)}=b_{3}V_{2}+b_{4}V_{4}-b_{1}V_{6},\text{ }%
X_{3}^{(I)}=p_{3}V_{2}+p_{4}V_{3}-p_{1}V_{5}  \label{3.22} \\
X_{4}^{(I)} &=&s_{4}V_{2}+s_{6}V_{3}+s_{7}V_{4}+s_{5}V_{5}-(s_{1}+s_{5})V_{6}
\label{3.23} \\
X_{5}^{(I)}
&=&m_{3}V_{1}+m_{4}V_{2}+m_{1}m_{3}V_{3}+2m_{2}m_{3}V_{4}-m_{3}V_{5}+2m_{3}V_{6}
\label{3.24} \\
X_{6}^{(I)} &=&\frac{n_{1}}{3}%
V_{1}+n_{4}V_{2}+n_{1}n_{2}V_{3}+n_{1}n_{3}V_{4}-\frac{n_{1}}{3}V_{5}-\frac{%
n_{1}}{3}V_{6}  \label{3.25}
\end{eqnarray}

The next objective of this section is to determine the invariant solutions
of the analyzed model, generated by the Lie operators (\ref{3.14})-(\ref%
{3.19}). The similarity reduction method \cite{BlumanK} will be applied.

Consequently, the operator $X_{1}^{(I)}$ from (\ref{3.14}) has the
characteristic equations:%
\begin{equation}
\frac{dt}{a_{1}}=\frac{dy}{0}=\frac{dx}{0}=\frac{du}{0}.  \label{5.1}
\end{equation}%
By integrating these equations, one obtains three invariants:%
\begin{equation}
I_{1}^{(1)}=x,\text{ }I_{2}^{(1)}=y,\text{ }I_{3}^{(1)}=u.  \label{5.2}
\end{equation}%
By designating the invariant $I_{3}^{(1)}$ $=H_{1}(x,y)$ as a function of
the other two and then inserting it to the Ricci flow equation (\ref{3.1}),
the similarity reduced equation takes the form:%
\begin{equation}
\left( H_{1}\right) _{x}\left( H_{1}\right) _{y}-H_{1}\left( H_{1}\right)
_{xy}=0.  \label{5.3}
\end{equation}%
The solution of this equation is:%
\begin{equation}
H_{1}(x,y)=h_{1}(x)h_{2}(y),\text{ }\forall \text{ }h_{1}(x),\text{ }\forall
h_{2}(y)  \label{5.4}
\end{equation}%
Thereby, the invariant solution corresponding to the operator $X_{1}^{(I)}$
has the final form:%
\begin{equation}
u_{1}^{(I)}(x,y)=h_{1}(x)h_{2}(y),\text{ }\forall \text{ }h_{1}(x),\text{ }%
\forall h_{2}(y)  \label{5.5}
\end{equation}%
\qquad \qquad Through a similar algorithm, the following results could be
obtained:

$(1)$ The Lie operators (\ref{3.15}), (\ref{3.16}), (\ref{3.17}) do
generate, respectively, the following similarity variables:%
\begin{eqnarray}
v_{2} &=&x,\text{ }w_{2}=\frac{b_{1}y-b_{4}}{b_{1}}\exp \left( \frac{b_{1}}{%
b_{3}}t\right) ;\text{ }v_{3}=y,\text{ }w_{3}=\frac{p_{1}x-p_{4}}{p_{1}}\exp
\left( \frac{p_{1}}{p_{3}}t\right)  \notag \\
v_{4} &=&\frac{s_{5}x+s_{6}}{s_{5}}\exp \left( -\frac{s_{5}}{s_{4}}t\right) ,%
\text{ }w_{5}=\frac{(s_{5}+s_{1})y-s_{7}}{s_{5}+s_{1}}\exp \left( \frac{%
s_{5}+s_{1}}{s_{4}}t\right)  \label{5.6}
\end{eqnarray}%
and, through similarity reduction, evolutionary equations will look like:%
\begin{equation*}
b_{1}w_{2}\left( H_{2}\right) ^{2}\left( H_{2}\right) _{w_{2}}+b_{1}\left(
H_{2}\right) ^{3}-
\end{equation*}%
\begin{equation}
-b_{3}H_{2}(H_{2})_{v_{2}w_{2}}+b_{3}\left( H_{2}\right) _{v_{2}}\left(
H_{2}\right) _{w_{2}}=0  \label{5.7}
\end{equation}%
\begin{equation*}
p_{1}w_{3}\left( H_{3}\right) ^{2}\left( H_{3}\right) _{w_{3}}+p_{1}\left(
H_{3}\right) ^{3}-
\end{equation*}%
\begin{equation}
-p_{3}H_{3}(H_{3})_{v_{3}w_{3}}+p_{3}\left( H_{3}\right) _{v_{3}}\left(
H_{3}\right) _{w_{3}}=0  \label{5.8}
\end{equation}%
\begin{equation*}
s_{5}v_{4}\left( H_{4}\right) ^{2}\left( H_{4}\right)
_{v_{4}}-(s_{5}+s_{1})w_{4}\left( H_{4}\right) ^{2}\left( H_{4}\right)
_{w_{4}}-
\end{equation*}%
\begin{equation}
-s_{1}\left( H_{4}\right) ^{3}+s_{4}H_{4}(H_{4})_{v_{4}w_{4}}-s_{4}\left(
H_{4}\right) _{v_{4}}\left( H_{4}\right) _{w_{4}}=0  \label{5.9}
\end{equation}

\bigskip $(2)$ The invariant solutions of the model, associated to $%
X_{2}^{(I)},$ $X_{3}^{(I)}$ respectively $X_{4}^{(I)},$ become stationary: 
\begin{eqnarray}
u_{2}^{(I)}(x,y) &=&\frac{b_{1}}{b_{1}y-b_{4}}F_{2}(x),\text{ }%
u_{3}^{(I)}(x,y)=\frac{m_{1}}{m_{1}x-m_{3}}G_{3}(y),\text{ }\forall \text{ }%
F_{2}(x),\text{ }\forall \text{ }G_{3}(y)  \label{5.101} \\
u_{4}^{(I)}(x,y) &=&q_{1}\left( \frac{s_{5}x+s_{6}}{s_{5}}\right)
^{q_{2}}\left( \frac{(s_{5}+s_{1})y-s_{7}}{s_{5}+s_{1}}\right) ^{\left(
s_{5}q_{2}-s_{1}\right) /(s_{5}+s_{1})},\text{ }\forall q_{1},q_{2}=const.
\label{5.102}
\end{eqnarray}%
$(3)$ By integrating characteristic equations, the Lie operators (\ref{3.18}%
) and (\ref{3.19}) do lead to the following new variables:%
\begin{eqnarray}
v_{5} &=&m_{4}x+m_{3}(x-m_{1})t,\text{ }w_{5}=\frac{y+m_{2}}{%
(m_{3}t+m_{4})^{2}}  \notag \\
v_{6} &=&x(n_{1}t+3n_{4})-n_{1}n_{2}t,\text{ }%
w_{6}=y(n_{1}t+3n_{4})-n_{1}n_{3}t  \label{5.11}
\end{eqnarray}%
Through the similarity reduction method, they generate the reduced
differential equations:%
\begin{equation*}
m_{3}(v_{5}-m_{1}m_{4})\left( H_{5}\right) ^{2}\left( H_{5}\right)
_{v_{5}}-2m_{3}w_{5}\left( H_{5}\right) ^{2}\left( H_{5}\right) _{w_{5}}-
\end{equation*}%
\begin{equation}
-H_{5}(H_{5})_{v_{5}w_{5}}+\left( H_{5}\right) _{v_{5}}\left( H_{5}\right)
_{w_{5}}=0  \label{5.12}
\end{equation}%
\begin{equation*}
n_{1}(v_{6}-3n_{2}n_{4})\left( H_{6}\right) ^{2}\left( H_{6}\right)
_{v_{6}}+n_{1}(w_{6}-3n_{3}n_{4})\left( H_{6}\right) ^{2}\left( H_{6}\right)
_{w_{6}}+
\end{equation*}%
\begin{equation}
+3n_{1}\left( H_{6}\right) ^{3}-H_{6}(H_{6})_{v_{6}w_{6}}+\left(
H_{6}\right) _{v_{6}}\left( H_{6}\right) _{w_{6}}=0  \label{5.13}
\end{equation}%
$(4)$ The invariant solutions of the model, corresponding to $X_{5}^{(I)},$
respectively to $X_{6}^{(I)}$are nonstationary. They have the expressions:%
\begin{eqnarray}
u_{5}^{(I)}(t,x,y) &=&\rho _{1}\left[ (m_{1}m_{4}-v_{5})w_{5}^{(1/2)}\right]
^{-\rho _{2}},\text{ }\forall \text{ }\rho _{1},\rho _{2}=const.  \notag \\
u_{6}^{(I)}(t,x,y) &=&\gamma _{1}(v_{6}-3n_{2}n_{4})^{\gamma
_{2}}(-w_{6}+3n_{3}n_{4})^{-(3+\gamma _{2})},\text{ }\forall \gamma
_{1},\gamma _{2}=const.  \label{5.14}
\end{eqnarray}%
where invariants $v_{5},$ $w_{5},$ $v_{6},$ $w_{6}$, expressed in the terms
of original variables, are provided by (\ref{5.11}).

\subsection{Symmetries and invariant solutions for Case (II)}

By solving the latter five equations of the determining system (\ref{3.13})
under the conditions (\ref{3.133}), the following Lie symmetry operators
will be generated : 
\begin{eqnarray}
X_{1}^{(II)} &=&c_{2}\frac{\partial }{\partial t}+\xi (x)\frac{\partial }{%
\partial x}-\left( \frac{d\xi (x)}{dx}u\right) \frac{\partial }{\partial u},%
\text{ }c_{2}=const.,\text{ }\forall \xi (x)  \label{6.1} \\
X_{2}^{(II)} &=&c_{2}^{\prime }\frac{\partial }{\partial t}+\eta (y)\frac{%
\partial }{\partial y}-\left( \frac{d\eta (y)}{dy}u\right) \frac{\partial }{%
\partial u},\text{ }c_{2}^{\prime }=const.\text{ }\forall \eta (y)
\label{6.2}
\end{eqnarray}%
\textit{Remark 2: }The forms (\ref{6.1}), (\ref{6.2}) represent particular
cases of the general operator (\ref{3.20}). They could be obtained by
imposing $c_{1}=0,$ $\eta (y)=0,$ respectively $c_{1}=0,$ $\xi (x)=0$ in (%
\ref{3.20}).

Let us obtain invariant solutions for the $2D$ Ricci model, associated to
operator (\ref{6.1}). Similar results will be derived for operator (\ref{6.2}%
).

The function $u=\Psi (t,x,y)$ is an invariant solution generated by (\ref%
{6.1}), providing that:%
\begin{equation}
X_{1}^{(II)}\left[ u-\Psi (t,x,y)\right] \left\vert _{u=\Psi }\right. =0
\label{6.3}
\end{equation}%
The previous condition has the equivalent form:%
\begin{equation}
\dot{\xi}(x)\Psi +c_{2}\Psi _{t}+\xi (x)\Psi _{x}=0  \label{6.4}
\end{equation}%
The general solution of (\ref{6.4}) is:%
\begin{equation}
\Psi (t,x,y)=\frac{F\left( v,w\right) }{\xi (x)}\text{ },\text{ }v=y,\text{ }%
w=t-\int \frac{c_{2}}{\xi (x)}dx  \label{6.6}
\end{equation}%
with arbitrary functions $\xi (x)$ and $F\left( v,w\right) .$

The substitution of (\ref{6.6}) into the $2D$ equation (\ref{3.1}) yields a
reduced differential equation of the form:%
\begin{equation}
F^{2}F_{w}+c_{2}FF_{vw}-c_{2}F_{v}F_{w}=0  \label{6.7}
\end{equation}%
which admits as a solution an arbitrary function $F_{1}(v).$

Consequently, by taking into account (\ref{6.6}), the invariant solution
associated to operator (\ref{6.1}) becomes:%
\begin{equation}
u_{1}^{(II)}(x,y)=\frac{F_{1}(y)}{\xi (x)},\text{ }\forall F_{1}(y),\text{ }%
\forall \xi (x)  \label{6.8}
\end{equation}%
Due to similar reasons, the invariant solution attached to operator (\ref%
{6.2}) will be:%
\begin{equation}
u_{2}^{(II)}(x,y)=\frac{F_{2}(x)}{\eta (y)},\text{ }\forall F_{2}(x),\text{ }%
\forall \eta (y)  \label{6.9}
\end{equation}

\section{Concluding remarks}

In literature the problem of constructing a conserved vector is generally
approached by means of the direct method of solving $D_{i}P^{i}=0,$ for a
given equation with no recourse to symmetry properties. It usually involves
ad hoc assumptions able to simplify the procedure. For\ a known symmetry
operator $X,$ determining the conserved vector $P=(P^{i}),$ $i=\overline{1,p}
$ whould become simpler if the relation (\ref{2.7}) which connects
symmetries and conservation laws, should be added to the conservation law $%
D_{i}P^{i}=0.$ Conversely, the special relation (\ref{2.7}) could also be
used to obtain the symmetry generators $X$ associated with a given conserved
vector $P.$ The latter approach is applied in this paper for the $2D$ Ricci
flow model.

The central objects, in the study of conservation laws, are the sets of
conservation law multipliers. The key property of these multipliers is that
their existence implies the existence of conservation laws. The main results
of this paper consist in obtaining: $(i)$\ the set of multipliers (\ref{3.7}%
) which depend up on two arbitrary functions $f(x)$ and $g(y)$; $(ii)$ the
set of local conservation laws (\ref{3.9}) associated to the mentioned
multipliers; $(iii)$ two sets of symmetry operators (\ref{3.14})-(\ref{3.19}%
) and (\ref{6.1}), (\ref{6.2}), by making use of the the tight relationship (%
\ref{2.7}) between symmetries and conservation laws and by solving the
determining differential system (\ref{3.13}) in two different cases. In the
first case, each of the Lie symmetry generators could be expressed as a
linear combination of the independent operators (\ref{3.21}), obtained \cite%
{Ricci} for the linear sector of invariance of the Ricci flow model. In the
second case, both operators (\ref{6.1}) and (\ref{6.2}), could be identified
as particular forms of the general Lie operator \cite{Ricci} which depends
on two arbitrary functions; $(iv)$ six stationary and two nonstationary
invariant solutions of the Ricci model, associated to the mentioned symmetry
operators are computed. Among, them, some new similarity solutions, not yet
given in literature, were highlighted. They are listed in (\ref{5.102}), (%
\ref{5.11}), (\ref{5.14}), (\ref{6.8}), (\ref{6.9}).

These results have been obtained by assuming a conservation law multiplier
which does not depend on the derivatives of the variable $U(t,x,y)$. The
multipliers which depend on the first order and higher order partial
derivatives of $U$ may lead to further conservation laws. They will be
studied in forthcoming works.

\textbf{Acknowledgements}

The author is grateful for the financial support offered by the Romanian
Ministry of Education, Research and Innovation, through the National Council
for Scientific Research in Higher Education (CNCSIS), in the frame of the
Programme "Ideas", grant code ID 418/2008.

\end{document}